\documentclass[aps,prl,noshowpacs,twocolumn,reprint,amsmath,amssymb,superscriptaddress,longtable]{revtex4-1}

\usepackage{graphicx}% Include figure files
\usepackage{dcolumn}% Align table columns on decimal point
\usepackage{bm}% bold math
\usepackage{tablefootnote}

\usepackage{amsthm}
\usepackage{mathtools}
\usepackage{amsmath}
\usepackage{amssymb}
\usepackage{graphicx}
\usepackage{url}
\usepackage{float}
\usepackage{bm}
\usepackage{ifthen}
\usepackage[usenames,dvipsnames]{color}
\usepackage{mathrsfs}
\usepackage[colorlinks=true,citecolor=blue,urlcolor=black]{hyperref}
\usepackage{float}
\usepackage{subfigure}
\usepackage{enumitem}
\usepackage{color}
\usepackage{setspace}
\usepackage{braket}
\usepackage{comment}
\usepackage{lipsum}
\usepackage{color}

\begin{document}

\title{ Passive  continuous variable quantum key distribution}
\author{Chenyang Li}
\email{Corresponding author email:licheny@hku.hk}
\affiliation{Department of Physics, University of Hong Kong, Pokfulam, Hong Kong}
\affiliation{ Department of Electrical \& Computer Engineering, University of Toronto, M5S 3G4, Canada}
\author{Chengqiu Hu}
\thanks{The author contributes equally as the first author}
\affiliation{Department of Physics, University of Hong Kong, Pokfulam, Hong Kong}

\author{Wenyuan Wang}
\affiliation{Department of Physics, University of Hong Kong, Pokfulam, Hong Kong}
\author{Rong Wang}
\affiliation{Department of Physics, University of Hong Kong, Pokfulam, Hong Kong}
\author{Hoi-Kwong Lo}
\affiliation{Department of Physics, University of Hong Kong, Pokfulam, Hong Kong}
\affiliation{ Department of Electrical \& Computer Engineering, University of Toronto, M5S 3G4, Canada}
\affiliation{Department of Physics, University of Toronto, Toronto, ON M5S 3G4, Canada}
\affiliation{Quantum Bridge Technologies, Inc., 100 College Street, Toronto, ON M5G 1L5, Canada.}

\begin{abstract}
Passive quantum key distribution (QKD) has been proposed for discrete variable (DV) protocols to eliminate side channels in the source.  Unfortunately, the key rate of passive DV-QKD protocols suffers from sifting loss and additional quantum errors.   In this work, we propose the general framework of passive continuous variable quantum key distribution. Rather surprisingly, we find that the passive source is a perfect candidate for the discrete-modulated continuous variable quantum key distribution (DMCV QKD) protocol. With the phase space remapping scheme, we show that passive  DMCV QKD offers the same key rate as its active counterpart. Considering the important advantage of removing side channels that have plagued the active ones, passive DMCV QKD is a promising alternative. In addition, our protocol makes the system much simpler by allowing modulator-free quantum key distribution. Finally, we experimentally characterize the passive DMCV QKD source, thus showing its practicality.
\end{abstract}

\date{\today}
\maketitle

\textit{Introduction.}
Quantum key distribution (QKD)\cite{bennet1984quantum,ekert1991quantum} allows two distant parties to establish information-theoretically secure
keys against any eavesdropper based on the laws of quantum mechanics \cite{lo2014secure}.
Both discrete-variable (DV) QKD protocols based on single photon detection \cite{lo2014secure,xu2020secure} and continuous-variable (CV) QKD protocols \cite{diamanti2015distributing,weedbrook2012gaussian} based on coherent detection have been demonstrated in practice.

During the implementation of  QKD protocols, practical devices in a QKD system might contain side channels or imperfections, which could compromise the security of QKD \cite{xu2020secure,diamanti2015distributing}. Without assumptions of the practical devices,  device-independent(DI) QKD protocol \cite{pironio2009device} is an idealized but challenging solution since the security of DI QKD is only guaranteed by the violation of Bell inequality. Nevertheless, the high demands for experimental implementations such as small losses, limit its wide applications. A more practical protocol, measurement-device-independent (MDI) QKD \cite{lo2012measurement} (see also \cite{braunstein2012side}) has
been proposed to eliminate all side channels in the measurement devices.  For the most state-of-the-art DV and CV  QKD  implementations,   the sources are still assumed to be trusted and the prepared states are assumed to be perfect.  However, in practice,
modulators can have side channels that possibly directly leak information to Eve and be vulnerable to Trojan Horse attacks\cite{gisin2006trojan,khan2014trojan}. Also, the perfect modulation can never be achieved in the state preparation due to the many factors, such as the resolution of modulation depth\cite{jouguet2012analysis}, the stability of modulators\cite{liu2017imperfect}, the correlations between adjacent pulses\cite{yoshino2018quantum}, the laser intensity flucatuation\cite{laudenbach2018continuous}  and etc. Many practical methods have been proposed to solve the individual imperfection and quantify the potentially leaked information\cite{li2021simple,huang2016long,derkach2017continuous,usenko2010feasibility,jacobsen2015continuous,derkach2016preventing,tamaki2016decoy,pereira2019quantum,pan2020practical,hajomer2022modulation}, but none of these methods can have a one-time solution to handle all these possible side channels or imperfections in the source. Therefore, fully-passive DV QKD \cite{wang2022fully} has been recently proposed to eliminate the side channels from the source  based on the previous passive DV protocols \cite{curty2009non,curty2010passive}. A passive QKD source means there is no active modulator and the encoding is based on the random measurement outcomes. In particular,  the encoding of passive  DV QKD is entirely performed via the post-section of a small region of detection results.  This post-selection will unavoidably cause sifting losses and the intensities of decoy states are not fixed so that it will also cause more quantum errors in the parameter estimations. Therefore, the key rate of passive DV QKD will be at least one order of magnitude less than the active counterpart\cite{wang2022fully}.
On the other hand, continuous variable quantum key distribution has the potential for high-key rates and low-cost implementations using current standard telecom
components\cite{diamanti2015distributing,weedbrook2012gaussian}. Previous passive continuous variable quantum key distributions are all based on the thermal source \cite{qi2020experimental,qi2018passive,huang2021experimental}, which is quite noisy compared to the coherent light sources.

In this letter, we propose the idea of passive continuous variable quantum key distribution based on coherent light sources, especially for the discrete-modulated continuous variable quantum key distribution (DMCV QKD), since  DMCV QKD is super interesting due to protocol simplicity and their great potential for massive deployment in the quantum-secured networks\cite{lin2019asymptotic,ghorai2019asymptotic,matsuura2021finite,liu2021homodyne,kaur2021asymptotic,denys2021explicit}. Recently, DMCV QKD has been experimentally demonstrated for  sub-Gbps key rates  within  the fiber \cite{roumestan2022experimental} and the metropolitan area\cite{wang2022sub}, showing the
path for future high-rate and large-scale CVQKD deployment in secure broadband metropolitan and access networks.
Here, we surprisingly  find that the passive source is a perfect candidate for DMCV QKD. That is, passive DMCV QKD will not have any additional sifting loss and quantum errors based on the phase space remapping scheme, and therefore have the exact same key rate with respect to the active modulated counterpart.  Moreover, we perform an experimental characterization of a source for passive DM-CV-QKD by testing its stability of intensity and the randomness of the phase. Furthermore, we analyze the resolution of phase and noise by comparison with the active modulation.  Finally, we look beyond the DMCV QKD and  propose the possible solution for the passive source for other CVQKD protocols such as  Gaussian modulation CVQKD.

\begin{figure}[!htb]
	\includegraphics[scale=0.5]{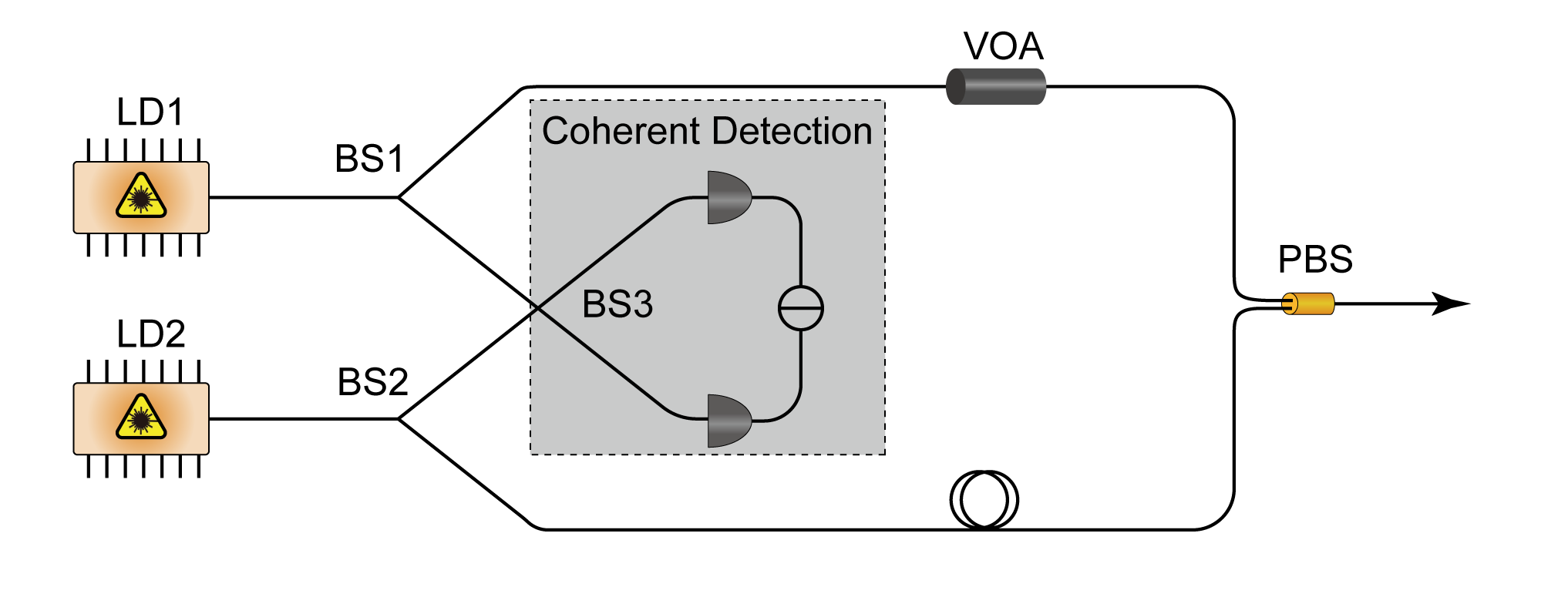}
	\caption{Passive source for DMCV QKD.  Two gain-switch lasers will generate coherent light pulses with random phases. One path will work as the signal, and another path will work as the phase reference, which will be privately held by Alice as encoding information. After time-and-polarization multiplexing, the signal and phase reference will be sent to Bob through a channel. BS: beamsplitter, PBS: polarization beamsplitter, VOA: variable optical attenuator.}
\end{figure}\label{fig:2}

\textit{Passive source}

A schematic passive source is illustrated in Fig.~1 and experimental implementation will be discussed later. The key point here is that the output pulses of the gain-switch laser can have fixed intensities with the same spectral characteristics, but totally \textbf{random} phases\cite{yuan2014robust}. Two laser diode are operated in the gain-switch mode, generating the phase-randomized coherent state pulses, i.e., $ |\sqrt{\mu_1}e^{i\phi_1}\rangle, |\sqrt{\mu_2}e^{i\phi_2}\rangle$, where $\mu_1$ and $\mu_2$ are predetermined, and $\phi_1$  and $\phi_2$ are uniformly distributed in the  $[0,2\pi)$. After BS1 and BS2, the two pulses will interfere at a beam splitter and then the phase difference $\theta=\phi_1- \phi_2$ is measured by coherent detection, where $\theta$ will also be uniformly distributed in the  $[0,2\pi)$. The first phase-randomized coherent state pulse, one output port of the BS1,  will be attenuated by a VOA to a predetermined intensity $\mu$ and then used as the signal. The second phase-randomized coherent state pulse, one output port of the BS2,  will work as the local oscillator or the phase reference of  the local local oscillator\cite{qi2015generating,soh2015self} depending on the intensity.  Here, we also propose the time-and-polarization multiplexing scheme for the first and second phase-randomized coherent state pulses by utilizing the delay line and Farader-mirror\cite{jouguet2013experimental,qi2007experimental}. After being combined by the PBS, the multiplexed pulses will be sent to Bob through an insecure channel.

With the phase reference, our signal state can be written as $|\alpha e^{i\theta}\rangle$, where $\alpha =\sqrt{\mu}$ is predetermined and $\theta$ will be uniformly distributed among  $[0,2\pi)$.
For DMCV QKD  protocol\cite{lin2019asymptotic}, Alice should encode four coherent states, i.e., $ |\alpha e^{i0}\rangle,|\alpha e^{i\pi/2}\rangle, |\alpha e^{i\pi}\rangle,|\alpha e^{i3\pi/2\rangle}$. To achieve this encoding,  one simple approach as DV protocols \cite{wang2022fully} is to post select a small slice around the four possible phases, such as$\{0\pm \Delta\phi ,\pi/2\pm \Delta\phi,\pi\pm \Delta\phi,3\pi/2\pm \Delta\phi\}$.  It is easy to show that this approach will have a high sifting loss and more quantum errors compared to the original DMCV QKD protocol, if  the mixed phases from $-\Delta\phi$  to $\Delta\phi$  are considered as an effect from a quantum channel.

\begin{figure}[!htb]
\centering
\includegraphics[scale=0.45]{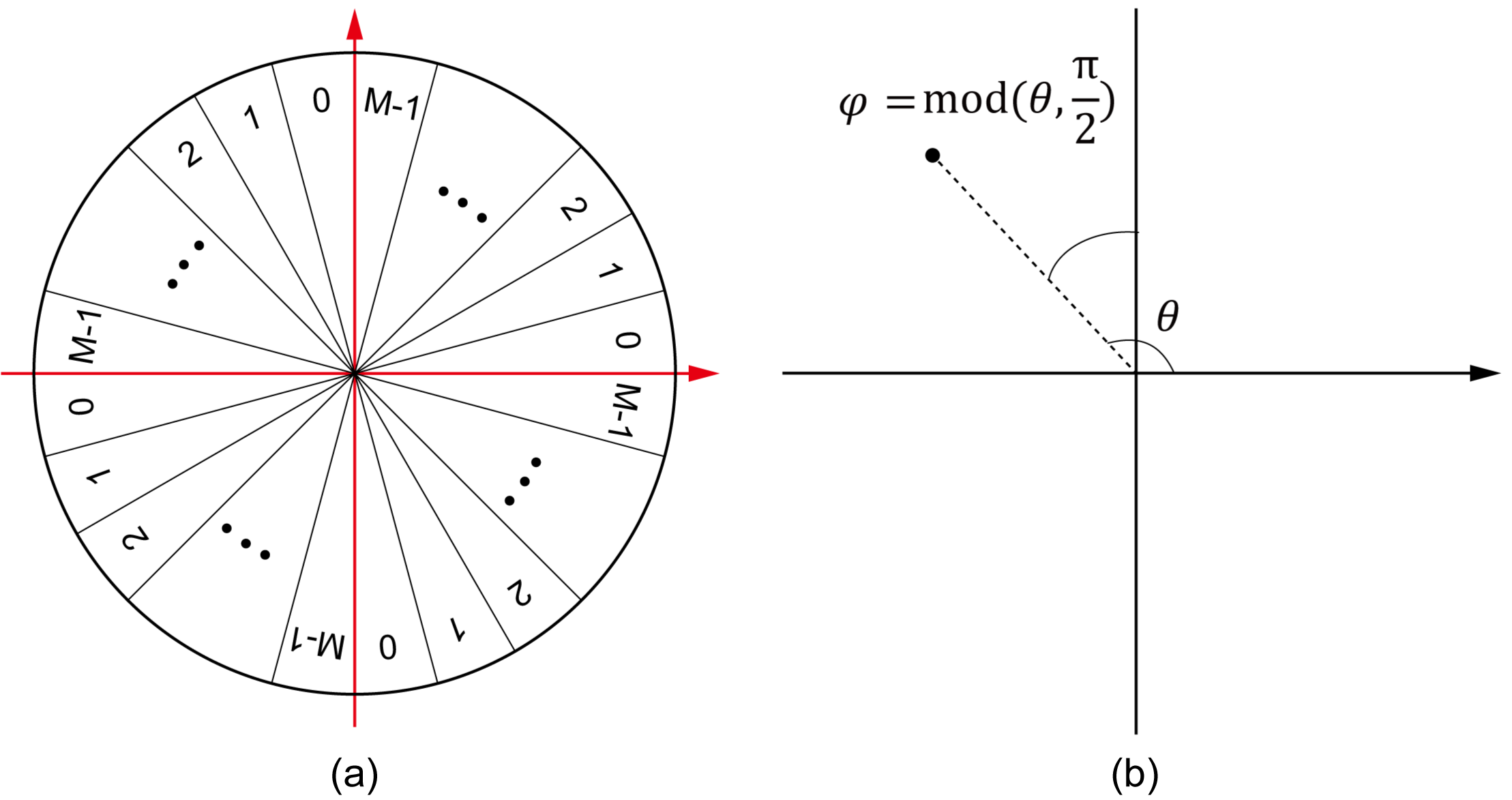}
\caption{Phase space remapping scheme. a) Alice divides the phase region into M slices and then tells Bob which slice the phase belongs to. This method will cause sifting loss and additional quantum errors. b) Alice only announce the remainder, $\text{mod}(\theta,\pi/2)$, to Bob. Bob can remap his phase space based on this additional information without any additional sifting loss and quantum errors.}
\end{figure}\label{idac}

Here comes the key point. To overcome all of these shortcomings, here, we propose a new and better method---phase space remapping scheme, which is also called quadrature remapping scheme in the previous CV QKD works\cite{qi2015generating,qi2007experimental}.  As depicted in Fig.~2(a), Alice divides the $[0,\pi/2)$ phase region into $M$ pieces of small slices, and groups every four slices of $\{\frac{n\pi}{2M}+0,\frac{n\pi}{2M}+\pi/2,\frac{n\pi}{2M}+\pi,\frac{n\pi}{2M}+3\pi/2\}$, where $n\in \{0,...,M-1\}$. For those groups with the same global phase basis information $\frac{n\pi}{2M}$, if Alice's measurement outcome $\theta$ satisfies  $ \frac{n\pi}{2M}\leq \text{mod}(\theta,\pi/2)< \frac{(n+1)\pi}{2M}$, where $\text{mod}(\theta,\pi/2)$ is the remainder of $\theta$ divided by $\pi/2$,  Alice can announce this global phase basis information $\frac{n\pi}{2M}$ so that Bob can realign the phase space. Meanwhile, Alice still secretly keeps the encoding information $x=\lfloor\theta,\pi/2\rfloor$, where $x$ is the quotient of $\theta$ divided by  $\pi/2$.  In this way, Alice can in principle divide the phases into infinitesimal slices, i.e, $M \rightarrow \infty$, and $\frac{n\pi}{2M} \rightarrow \text{mod}(\theta,\pi/2)$. As shown in Fig.~2(b), after Alice measures $\theta$, Alice will directly announce the global phase basis information $\varphi=\text{mod}(\theta,\pi/2)$ to Bob and therefore move all slices into the perfect encoding of $\{0,\pi/2,\pi,3\pi/2\}$.
Since all signals can be post-processed together, our passive source does not suffer from any finite-size penalty and sifting loss. By achieving the perfect modulation of four phases, the quantum errors will not be increased in our approach as well.

\begin{figure*}[!htb]
\centering
	\includegraphics[width=0.95\linewidth]{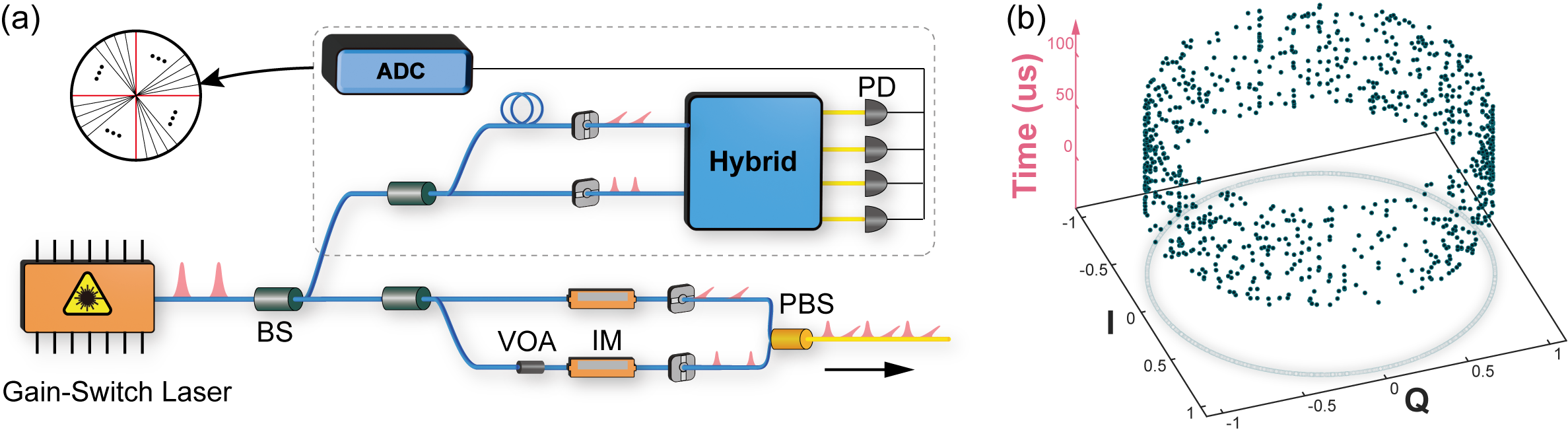}
	\caption{ a) Experimental setup of the passive source. A Gain-switch semiconductor laser is used to generate phase randomized pulses with a repetition rate of 20MHz and pulse width of 20ns. We use a 50/50 BS to separate the laser pulses into two parts, one for local measurement, and the other for distribution to Bob. For local measurement framed by the dashed box, the local oscillator and signal pulses are injected into a hybrid for coherent detection after being separated by a BS (the LO pulses go through the path with a 50ns delay). The pulses sent to Bob are also separated into two paths, one for LO pulses (with odd pulses blocked) and the other for signals (with even pulses blocked). A PBS is used to combine the LO and signal pulses so that they are multiplexed both in time and polarization.  IM: intensity modulator; PD: photodiode. (b) 3D scatter diagram showing the results of the local measurement of the passive source. Each dot in the plot denotes a measurement result at the corresponding time point and their projections are also shown on the IQ plane. }
\end{figure*}\label{fig:2}

\textit{Protocol description}
Our protocol is mainly based on the DMCV QKD with heterodyne detection in the reference\cite{lin2019asymptotic}. The main differences from protocols \cite{lin2019asymptotic} exist in the step of state preparation and post processing.  Note that our passive source can also fit for other DMCV QKD protocols\cite{ghorai2019asymptotic} with minor adjustments.

(1) State preparation. For each round $k\in N$, Alice obtains the phase information $\theta_k$ by coherent detection.   That is Alice prepares a coherent state $|\alpha e^{i\theta_k}\rangle$, where $\alpha$ is predetermined. Alice will privately hold the value $x_k=\lfloor\theta_k,\pi/2\rfloor$ as  a string where $x_k \in \{0,1,2,3\}$ and later publicly reveal the value $\varphi_k=\text{mod}(\theta_k,\pi/2)$ to Bob. Alice then sends signals and phase references to Bob through an insecure quantum channel.

(2)Measurement. After receiving Alice's state, Bob performs a heterodyne measurement on the state, which can be described by the positive operator-valued measure (POVM)
 $\{E_{\gamma}=(1/\pi)|\gamma\rangle\langle \gamma| \gamma\in C\} $.After applying this POVM, he obtains the measurement
outcome $y'_k\in C$.

(3)Post processing.  Bob obtains his final strings $y_k$ by applying the rotation matrix $M_{\varphi_k}$ (corresponding to  $\varphi_k$) to the measurement outcomes, that is $y_k=M_{\varphi_k}y'_k$.

The remaining steps of the protocol are standard, namely parameter estimation, information reconciliation, and privacy amplification.

\textit{Security analysis}
Here, we prove the security of the passive DMCV QKD protocol based on the equivalency of the passive and active sources.
Compared to the DMCV QKD  protocol in \cite{lin2019asymptotic},  Alice keeps the secret key information of $x_k=\lfloor\theta_k,\pi/2\rfloor$ and only announces the global phase basis information $\varphi_k=\text{mod}(\theta_k,\pi/2)$. Bob will do a phase rotation operation for all his measurement outcomes. This phase space remapping scheme is commonly used to compensate for phase drift in the CV QKD system and it will not affect the leaked information to Eve \cite{qi2015generating,qi2007experimental}.
As introduced before, this phase space remapping scheme can result in a perfect encoding of $\{0,\pi/2,\pi,3\pi/2\}$ without any additional sacrifices. In other words, the passive-encoding source is perfectly equivalent to the active-encoding source in the DMCV QKD protocol. Therefore, in principle, the key rate of the passive-encoding DMCV QKD should be identical to the active counterpart, while the passive DMCV QKD can be immune to all modulation imperfections and side channels. Furthermore, considering the resolution of coherent detection of strong light and the depth of the state-of-art modulation technology, the passive protocol will not sacrifice any secret key rate.

\textit{Experimental characterization of passive source}
We also experimentally characterize the encoder in a passive DMCV QKD setup to further verify the feasibility of our protocol. Gain-Switch semiconductor lasers are widely used in high-speed QKD and quantum random number generation\cite{yuan2014robust}. It can be easily modulated by electrical signals with a high repetition rate of up to GHz, of which the required driving voltage is much lower than the half-wave voltage needed by an electric-optical modulator. Moreover, a gain-switch laser can generate pulses with random phases, of which the randomness is inherited from spontaneous emission\cite{yuan2016directly, comandar2016quantum}. Here, we choose a 1550nm laser diode modeled EP1550-0-NLW as the source, utilizing the random features of the gain-switching process to demonstrate passive encoding. Fig.~3(a) shows the experimental setup of the passive source of DMCV QKD.  We drive the laser diode to generate 20MHz pulses with a pulse width of 20ns. The pulse sequences can be classified into two groups, ie. the odd-pulse group, and the even-pulse group, which are used as the local oscillators (LO) and the signals respectively. The phase difference between the LO and the signal pulse will be used as passive encoding information. A 50/50 beam splitter is used here to separate the pulses into two parts, one for local measurement (dashed box in Fig.~3(a)) and the other for distribution to Bob.

For the local measurement part, we further separate the pulses into two paths. On one of the paths, a 50ns fiber delay is added to align the LO and the signal pulses. Then the two paths are sent into the optical hybrid (from Optoplex company, more details see supplementary part \uppercase\expandafter{\romannumeral3}) for coherent detection. The local measurement results are recorded by Alice for raw keys generation as well as future global phase basis announcements. For the other part that will be sent to Bob, we also separate them into two paths, but different from the local part, there is no extra delay in either path. Instead, we use an intensity modulator with an extinction ratio of 30dB in both paths as an optical switch to block the odd (even) pulses. Notice that the modulators here are only used for blocking, rather than modulating, the encoding process is still fully passive. A variable optical attenuator (VOA) with about 70dB loss is used here to attenuate the signal to the required level. Then the two paths are combined by a PBS, after which the LO and signal are multiplexed both in time and polarization and ready for distribution. Fig.~3(b) shows the recorded results of the local measurement, from which we can see that the signal states are uniformly distributed on the unit circle in the constellation plane. Meanwhile, their phase values are randomly changed over time, which will be further processed by Alice and used for encoding and the global phase basis announcements  (see supplementary part \uppercase\expandafter{\romannumeral3}). The photodiodes (PDs) we used here are with high bandwidth of 5GHz. The ADC with a 5GSa/s sampling rate and 8-bit vertical resolution is used to record the output of the PDs. In addition, we also verify the output stability of the gain-switch laser (See supplementary part \uppercase\expandafter{\romannumeral3} ).

\textit{Resolution of phase and  noise analysis }
Here, we will discuss the resolution and noise in the passive source and make a  comparison with the active modulation.
With practical devices, the slice number $M$ can not be infinitely large.  Therefore, we consider practical devices such as the 8-bits ADC in the detection of the passive source and the 8-bits (modulation depth) modulator of active sources. As shown in supplementary part \uppercase\expandafter{\romannumeral1}, the resolution of phase in the passive sources $ \triangle\theta_1=\sqrt{2}/128=0.011$rad and the maximum fluctuation of phase in the active sources will be $ \triangle\theta_2=|\overline{\theta}-\theta|=2\pi/256=0.0245$rad. According to \cite{jouguet2012analysis}, both resolutions are good enough for the state preparation with extremely small errors.

Next, let us consider the noise in active sources and passive sources. In active sources, there can exist relative intensity noise coming from the laser and modulation noise from the DAC and modulator\cite{laudenbach2018continuous}. In the passive sources, as shown in the supplementary part \uppercase\expandafter{\romannumeral2},  the noise can be shown as
\begin{equation}
 \varepsilon=t_0\frac{|\mu_s|^2(2+E_{aq}-t_0t_{aq})}{|\mu_s|^2t_{aq}+4+4E_{aq}}.
\end{equation}\label{fig:2}
where $E_{aq}$ is the excess noise from the detector,  $t_0$  is the transmittance of the VOA, and $t_{aq}$ is Alice's  detection efficiency of Q quadrature, $|\mu_s|$ is the intensity of the classical coherent light.  It is expected that when $t_0<<1$, $\varepsilon$ approaches 0. For example, when the VOA has a 70dB attenuation coefficient, the $\varepsilon$ is roughly around $2\times10^{-7}$. Fig.~4 shows the  simulation results of the key rates over distance. From the right to the left lines, the VOA has a attenuation coefficient of 70dB, 30dB and 20 dB. The ideal case means there is no excess noise due to modulation, i.e., the VOA has a infinite large attenuation. Here, we find that, when the VOA have  a attenuation of more than 30dB, the excess noise of the
passive state preparation scheme can be effectively suppressed  and the key rate will almost be the same as the idea case.

\begin{figure}[!htb]
\centering
	\includegraphics[scale=0.5]{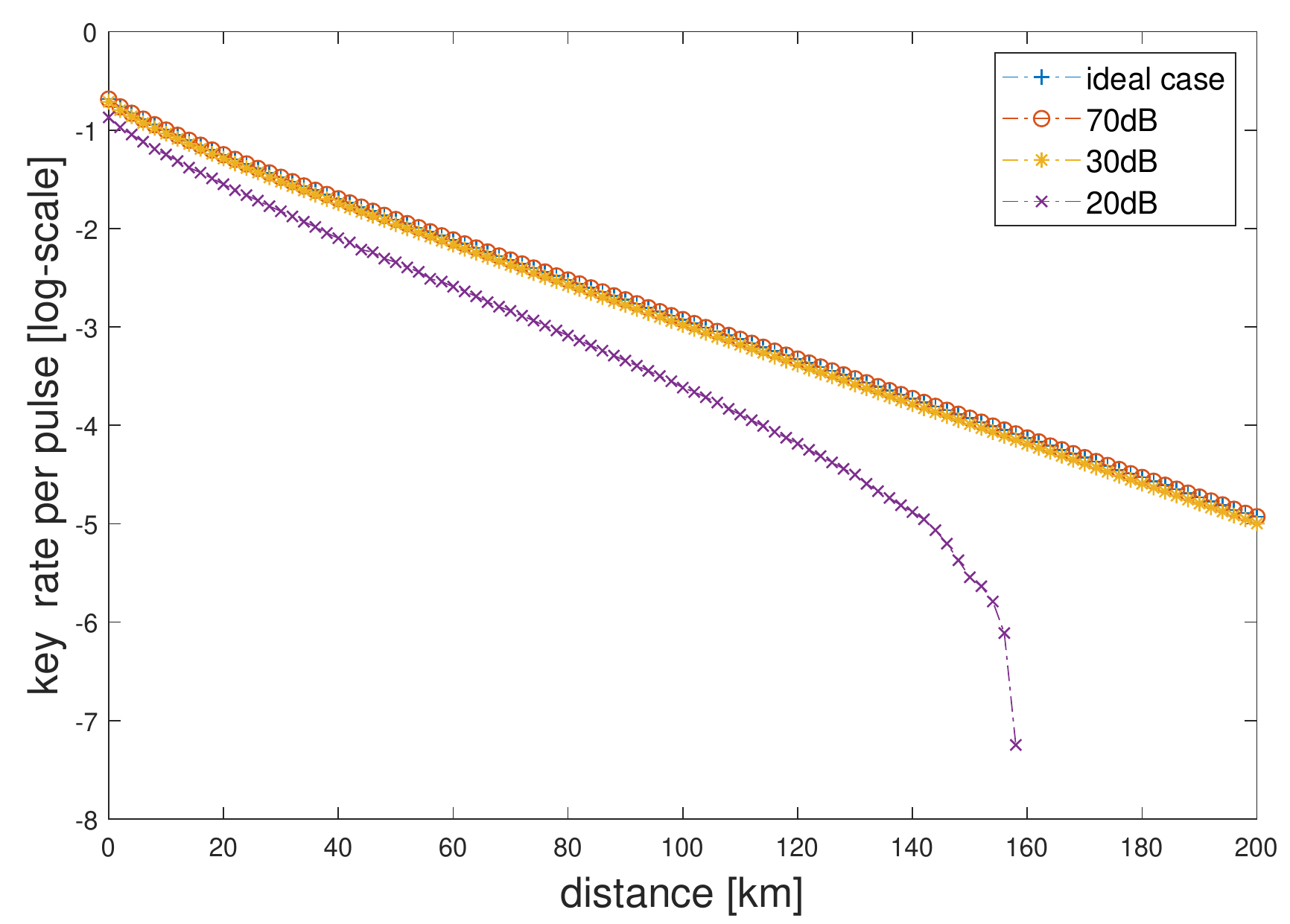}
	\caption{ Key rate versus distance. Here, we present the key rates  while the VOA attenuation coefficient differs. The ideal case means there is no excess noise due to the passive modulation. Here, the reconciliation efficiency is 0.95. Channel loss is 0.2dB/km. The coherent state amplitude is optimized by program. }
\end{figure}\label{}

\textit{Conclusion}
Here, we propose the general framework of passive continuous variable quantum key distribution based on the gain-switch laser sources. By the phase space remapping scheme we have proposed here,  passive DMCV QKD will not have any additional sifting loss and quantum errors, and therefore have the exact same key rate as its active counterpart.  Finally, we experimentally characterize the passive source for DMCV QKD, testing the randomness of the phase and the stability of the laser intensity.

Looking beyond DMCV QKD,  we can also build the passive source for other CV QKD protocols such as  Gaussian modulation(GM) CVQKD or quadrature amplitude modulation(QAM) CVQKD.  The main difference between the sources is the degree of freedom Alice has in
her output states. In the passive DMCV QKD sources we present here, only one degree of freedom is needed for phase modulation, which largely simplifies the complexity of the setup of the source. In other CV QKD protocols, at least two degrees of freedom will be needed, i.e., amplitude and phase modulation. Thus, we need multiple interferences to generate  two degrees of freedom. Also, we should  use the post-selection scheme described in the \cite{wang2022fully} to match the correct distribution of amplitude and phase, which will unavoidably cause  sifting losses and increase quantum errors. Analysis of the security of passive sources for GM CVQKD and QAM CVQKD will be presented in our future work.

\begin{acknowledgments} 
We acknowledge the financial support from the Natural Sciences and Engineering
Research Council of Canada (NSERC). We also acknowledge the funding from the University of Hong Kong start-up grant.
We also thank the Open QKD Security simulation programs.
\end{acknowledgments}

\end{document}

\title{Supplementary Material }

\title{ Passive  continuous variable quantum Key Distribution}
\author{Chenyang Li}
\email{licheny@hku.hk}
\affiliation{Department of Physics, University of Hong Kong, Pokfulam, Hong Kong}
\affiliation{ Department of Electrical \& Computer Engineering, University of Toronto, M5S 3G4, Canada}
\author{Chengqiu Hu}
\thanks{The author contributes equally as the first author}
\author{Wenyuan Wang}
\affiliation{Department of Physics, University of Hong Kong, Pokfulam, Hong Kong}
\author{Rong Wang}
\affiliation{Department of Physics, University of Hong Kong, Pokfulam, Hong Kong}
\author{Hoi-Kwong Lo}
\affiliation{Department of Physics, University of Hong Kong, Pokfulam, Hong Kong}
\affiliation{ Department of Electrical \& Computer Engineering, University of Toronto, M5S 3G4, Canada}
\affiliation{Department of Physics, University of Toronto, Toronto, ON M5S 3G4, Canada}
\affiliation{Quantum Bridge Technologies, Inc., 100 College Street, Toronto, ON M5G 1L5, Canada.}

\begin{abstract}
In this appendix, we will show the details of phase resolution and noise analysis of the passive source in DMCV QKD. Next, we give more details about the experimental setup and show the stability of intensity and the randomness of the phase.
 \end{abstract}
\maketitle
%----------------------------------------------------------------------------------------
%	SECTION 1
%----------------------------------------------------------------------------------------

\section{Resolution}

Here, we will discuss the resolution in the passive source and make a fair comparison with the active modulation.
 We consider practical devices such as the 8-bits ADC in the detection of the passive source and 8-bits in the modulation of the active sources.
As depicted in Fig 1.a, in the passive source, we will use the 8-bits to describe both positive and negative quadratures values. Therefore,  the phase resolution in the passive sources is $ \triangle\theta_1=\sqrt{2}/128=0.011$. As depicted in the Fig 1.b,  in the active sources,  8-bits will be used to describe the total $2\pi$. Therefore, the
 phase  resolution in the active sources will be $ \triangle\theta_2=2\pi/256=0.0245$.
According to \cite{jouguet2012analysis}, both resolution is good enough for the state preparation with extremely small errors.
\begin{figure}[htb]
\centering
\subfigure[ ]{
 \includegraphics [scale=0.7]{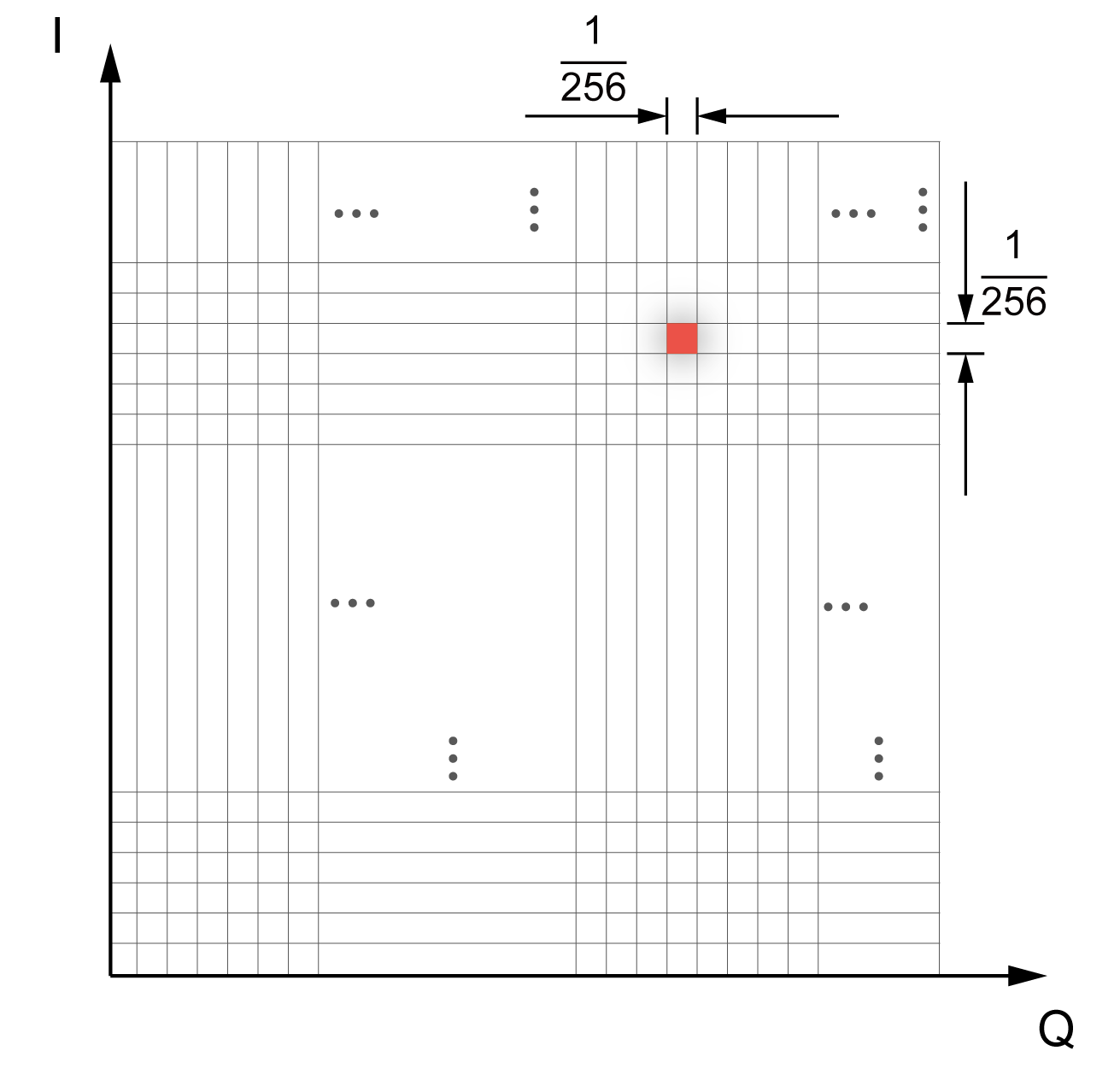}}
\subfigure[ ]{
\includegraphics [scale=0.7]{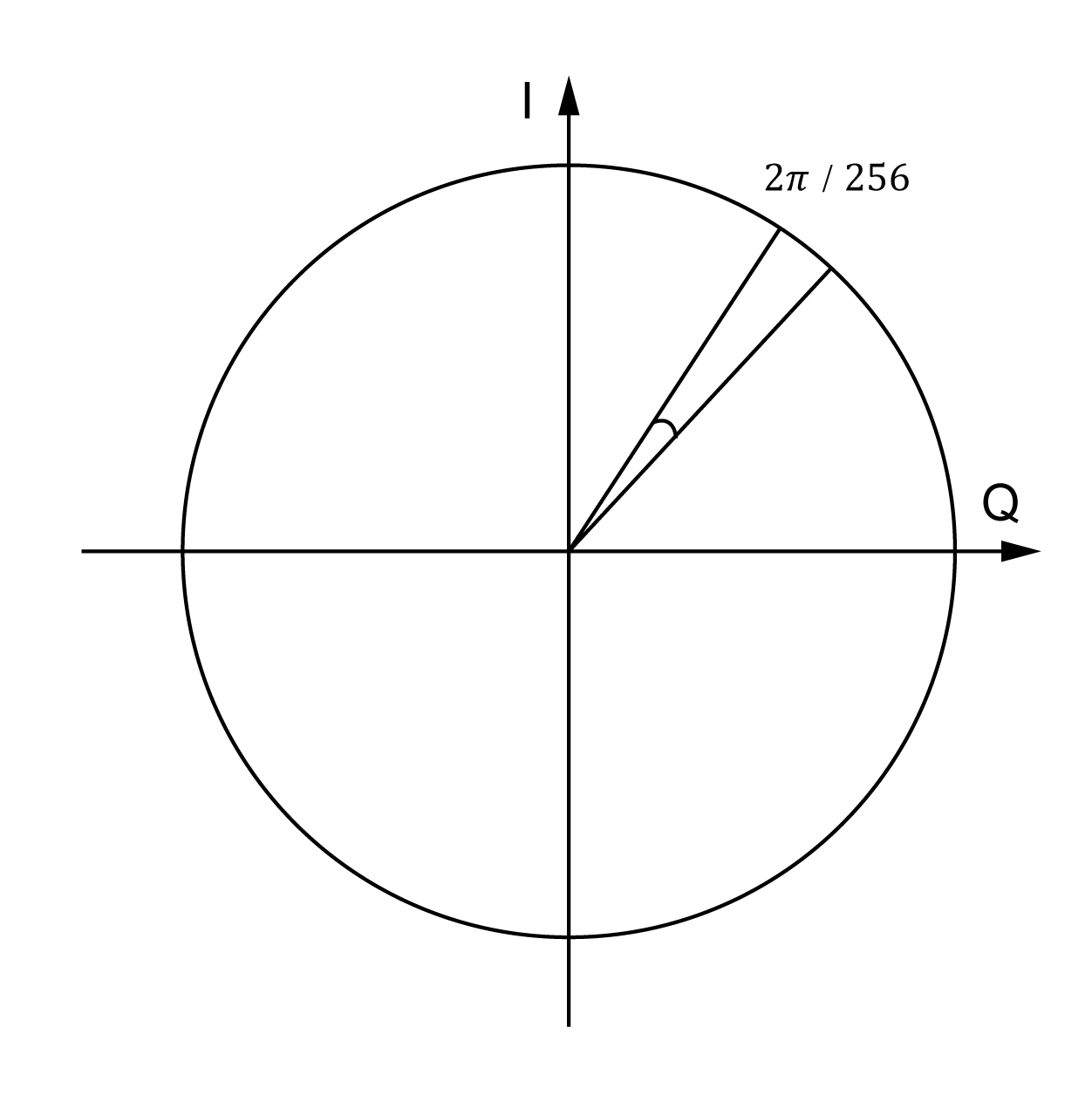}}

\caption{Phase resolution analysis }
\end{figure}\label{Fig4}

\section{Noise analysis in the passive CVQKD}

\begin{figure}[htb!]
\includegraphics [scale=0.7]{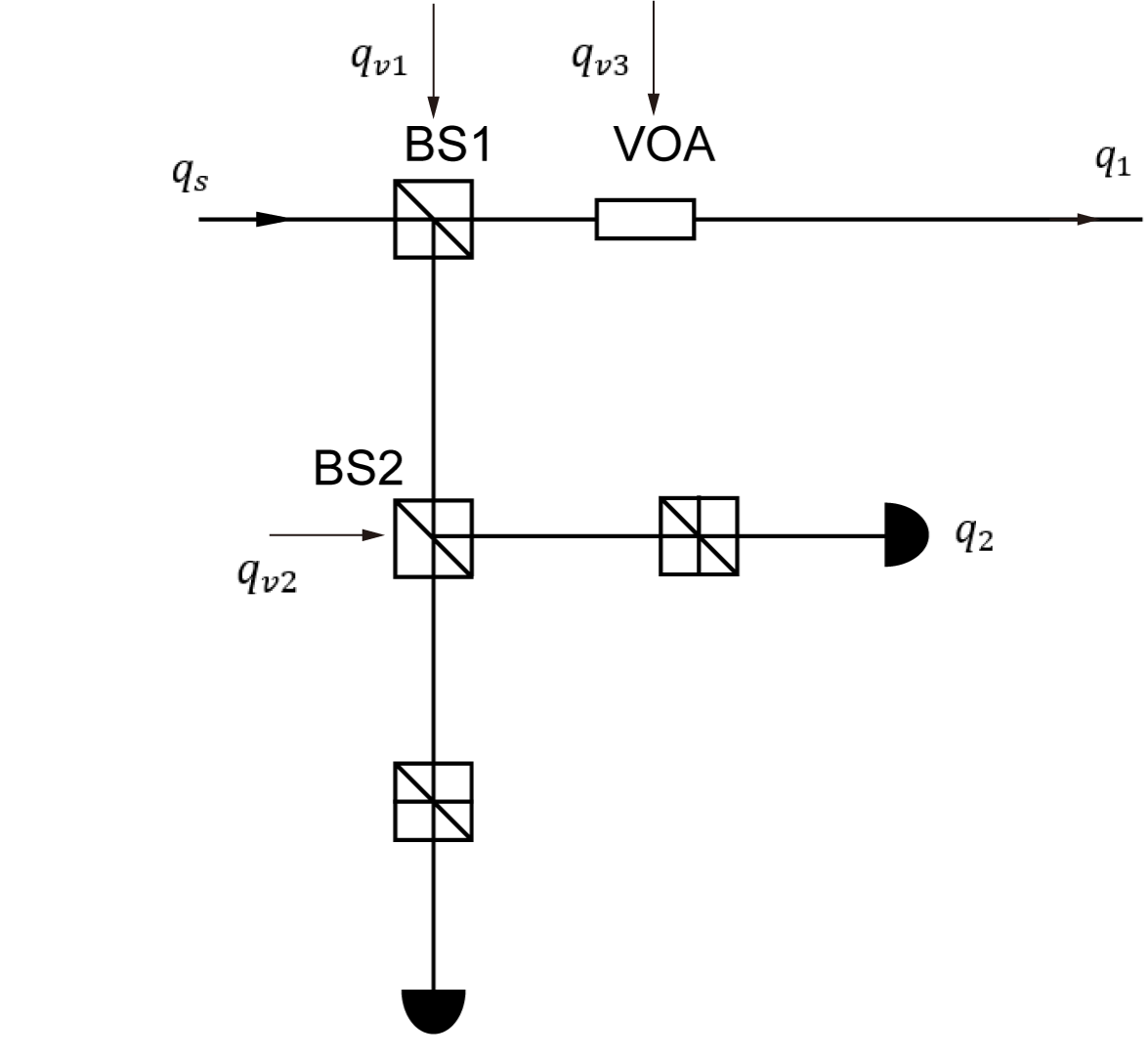}
\caption{ Passive source in  DMCV QKD }
\end{figure}\label{Fig2}
Here, we will discuss  the  noise in the passive  source as described in  Fig 2.  The main analysis is almost the same as \cite{qi2020experimental,qi2018passive,huang2021experimental}. The difference is that our source is coherent state pulses, which have a small quadrature variance and then behave less noisy.
For simplicity, we only consider the Q-quadrature below,  and the I-quadrature can be studied in a similar way. Based on the input mode $q_s$, the output mode can be described as:
\begin{equation}
q_1=\sqrt{\frac{t_0}{2}}q_s-\sqrt{\frac{t_0}{2}}q_{v1}-\sqrt{1-t_0}q_{v3},
\end{equation}
\begin{equation}
q_2=\frac{\sqrt{t_{aq}}}{2}q_s+\frac{\sqrt{t_{aq}}}{2}q_{v1}+\sqrt{\frac{t_{aq}}{2}}q_{v2}-\sqrt{1-t_{aq}}q_{aq}+e_{aq},
\end{equation}
where $q_s$ is the Q-quadrature of the output of the source,  $q_{v1},q_{v2},q_{v3}, q_{aq}$ are, separately, the quadrature of the vacuum state for the beam splitter.
For simplicity, we define  the relation $<(\triangle q_s)^2>= <(\triangle q_{vac})^2>=1$.
$t_0$  is the transmittance of the VOA, and $t_{aq}$ is the Alice's  detection efficiency of Q quadrature of coherent light. $e_{aq}$ represents the excess noise term from the homodyne detector with a noise variance $ E_{aq}= <(\triangle e_{aq})^2>.$  With phase space remapping scheme,  we can consider the source as $|\mu_s e^{i0}\rangle$ without losing generality. Then we have $ <q_s>^2=|\mu_s|^2, <q^2_s>=1+|\mu_s|^2$.

The excess noise of the passive source is defined as
\begin{equation}
\varepsilon=\frac{<(\triangle q_1)^2>}{<(\triangle q_{vac})^2>}-1 =<(\triangle q_1)^2>-1.
\end{equation}\label{fig:}
Give $q_2$, Alice's optimal estimation of $q_1$ is $\beta q_2$  where $\beta=<q_1q_2>/<q_2^2>$. We can determine that

\begin{equation}
\beta=\frac{|\mu_s|^2\sqrt{2t_0t_{aq}}}{|\mu_s|^2t_{aq}+4+4E_{aq}}.
\end{equation}

Alice's uncertainty on $q_1$ given her measurement result $q_2$ is $<(\triangle q_1)^2>=V_{q1|q2}=<(q_1-\beta q_2)^2>$.  Using previous equations, we can determine the excess noise of state preparation as
\begin{equation}
 \varepsilon=t_0\frac{|\mu_s|^2(2+E_{aq}-t_0t_{aq})}{|\mu_s|^2t_{aq}+4+4E_{aq}}.
\end{equation}\label{fig:2}

Equation (5)  suggests that  the excess noise of the
passive source can be effectively suppressed by introducing large optical attenuation at Alice's VOA. Also, since we detect the phase of classical light, $|\mu_s|^2$ will be the dominant term. Also, $t_0<<1$, $E_{aq}<<2$,  equation (5) can simplified as
\begin{equation}
 \varepsilon=\frac{2t_0}{t_{aq}},
\end{equation}\label{fig:2}
where $t_0$  is the transmittance of the VOA, and $t_{aq}$ is the Alice's  detection efficiency of Q quadrature of coherent light. With high detection efficiency  of classical coherent light, the excess noise approaches $2t_0$.

\section{Experimental details}
Here, we use a $90^{\circ}$ optical hybrid to implement the local measurement shown in Fig.3(a) in the main text. Fig.3 shows the diagram of the device. There are 2 input ports for LO and signal respectively. After going through the beamsplitters, LO and signal are both equally split into two paths. One of the LO paths will go through an extra $90^{\circ}$ phase change. The four outputs are denoted as $I_{1}$,$I_{2}$,$Q_{1}$ and $Q_{2}$, of which $I_{1}-I_{2}\propto sin\theta$ and $Q_{1}-Q_{2}\propto cos\theta$. Using four detectors after the outputs of the hybrid, we can calculate the phase difference between LO and signal pulses. The loss difference between LO and signals going through the device is less than 0.1dB.

\begin{figure}[h]
\includegraphics[scale=0.8]{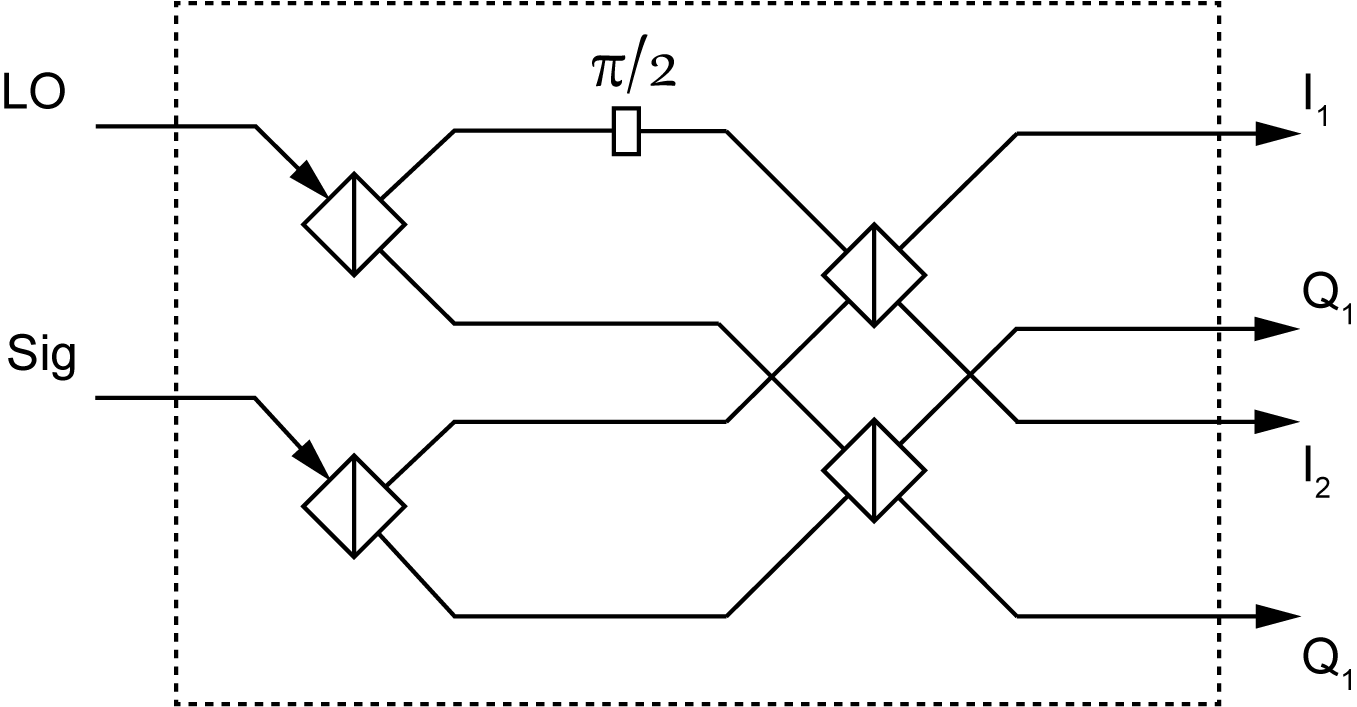}
\caption{$90^{\circ}$ optical hybrid}
\label{Fig3}
\end{figure}

To verify the feasibility of our protocol, we experimentally characterized the passive source by measuring the phase difference between LO and signal pulses. Fig.4 shows the distribution of the phase value of signals compared with LO pulses measured over time of $500\mu s$. The top half is the statistical histogram, showing the phase values are uniformly distributed between $[0,2\pi)$. The bottom half shows the randomly located phase scatters over time.

\begin{figure}[h]
\includegraphics[scale=0.6]{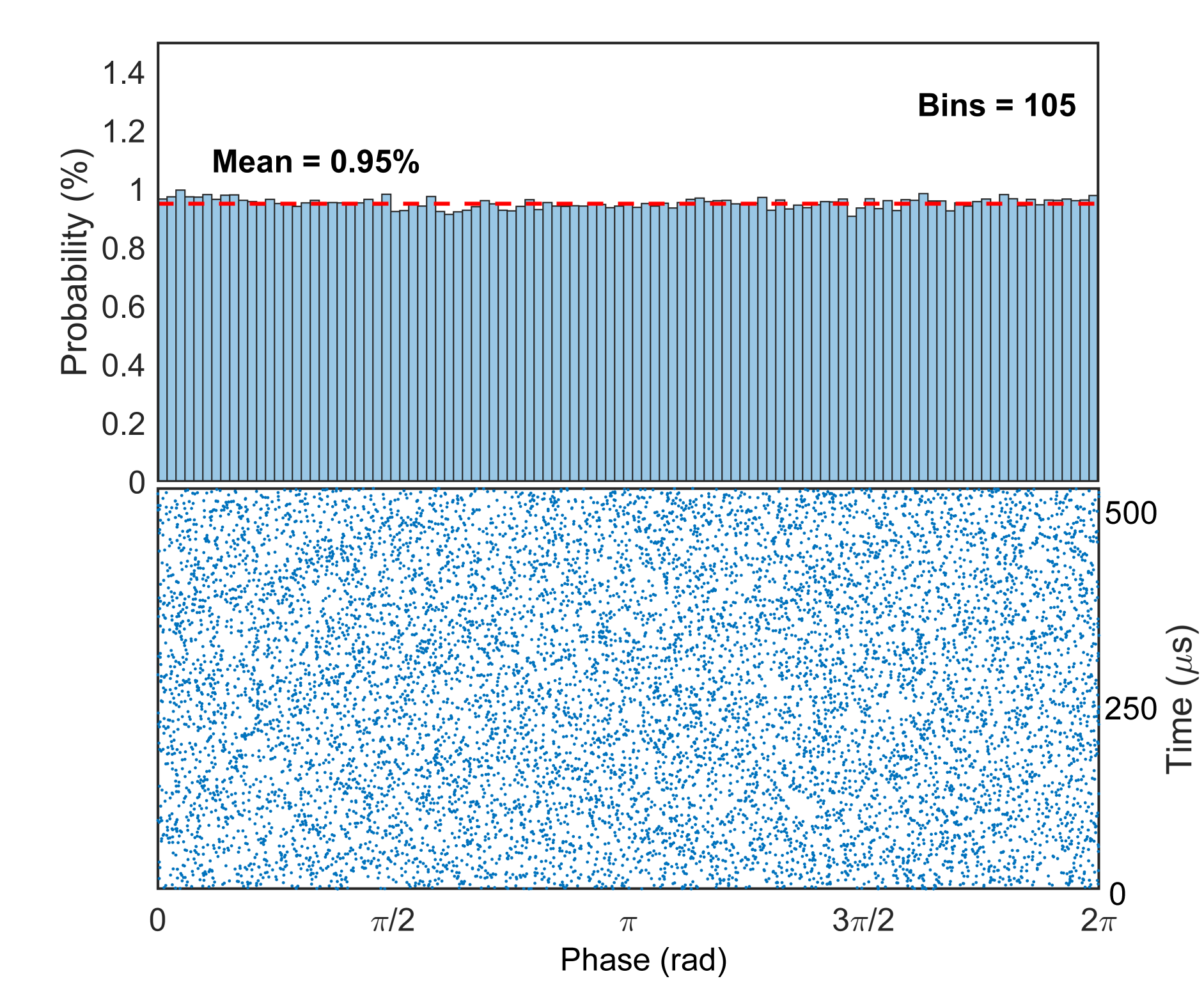}
\caption{Distribution of the measured phase values.}
\label{Fig4}
\end{figure}

To further show the randomness of the phases between pulses, we also calculate the autocorrelations of the phase sequence using a data size of $2\times 10^6$. By calculating the correlation between the delayed sequences and the original ones, we obtain the autocorrelations shown in Fig.5. In theory, for a truly random sequence with a data size of $10^{7}$, the average value of autocorrelation is 0 with a standard deviation of $4\times 10^{-4}$. \cite{ma2013postprocessing} The average residual value of our measured autocorrelations is $1.2\times 10^{-4}$, which shows true randomness.

\begin{figure}[h]
\includegraphics[scale=0.7]{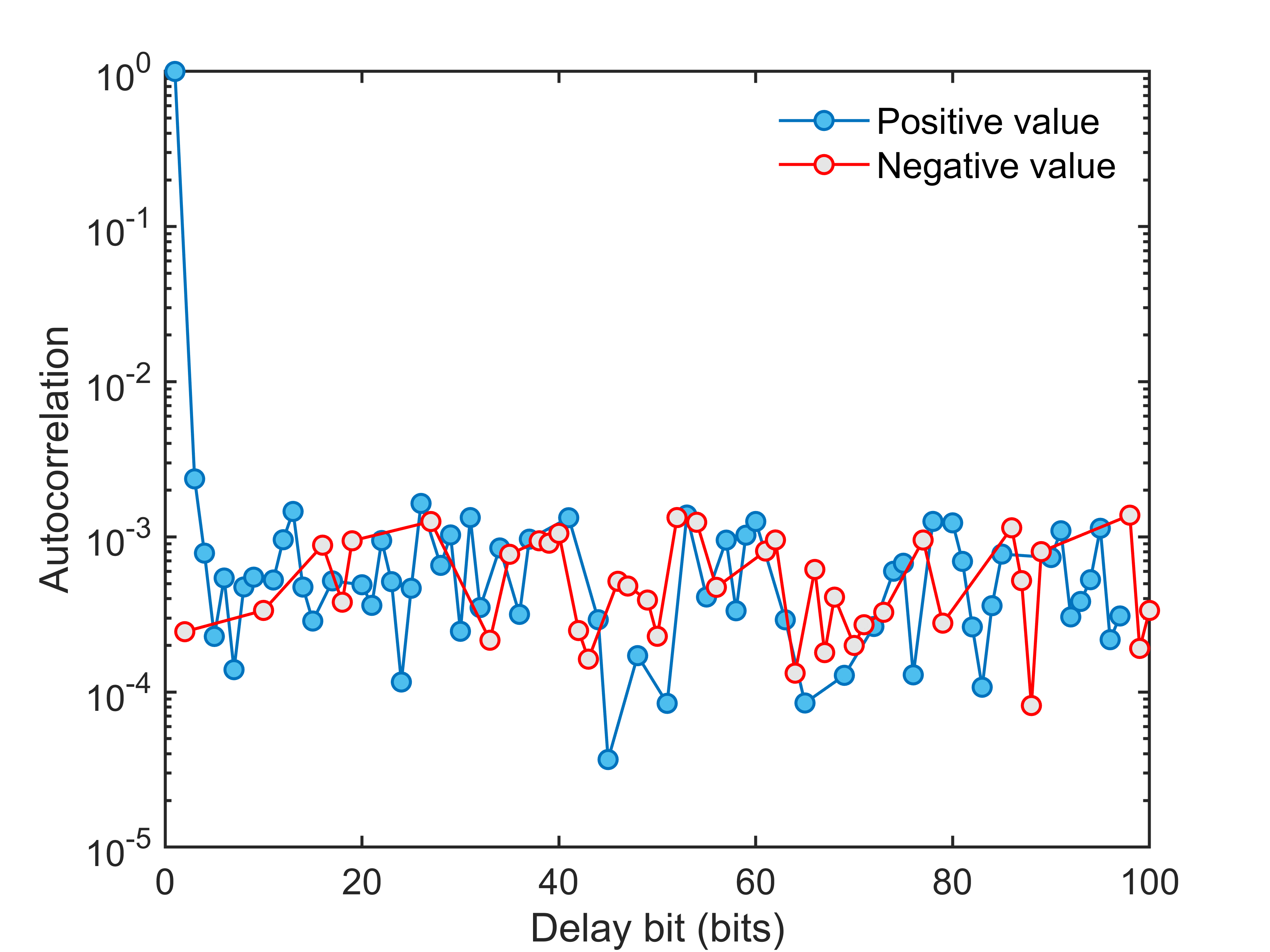}
\caption{Autocorrelation of measured phase values. The red dots represent negative values and the blue ones represent positive values. The average residual value is $1.2\times 10^{-4}$.}
\label{Fig5}
\end{figure}

In addition, we also test the output stability of the gain-switch laser. As shown in Fig. 6, the output intensity is $0.991\pm0.0024$. The standard deviation corresponds to 0.24\% of the mean value.

\begin{figure}[!htb]
	\includegraphics[scale=0.45]{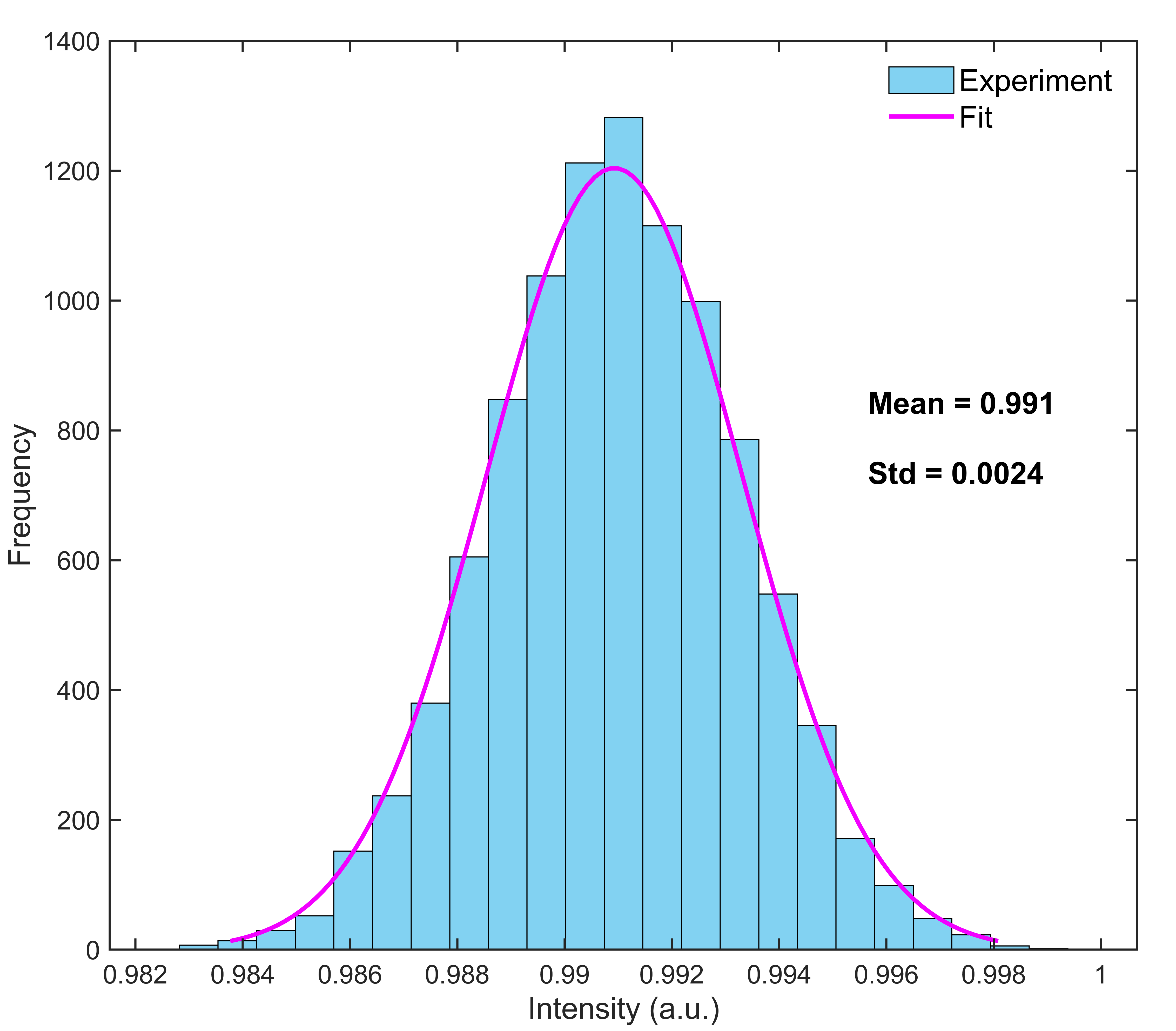}
	\caption{Insentity of the output pulse from Gain-switch laser. We test that the output intensity is $0.991\pm0.0024$. The standard deviation corresponds to 0.24\% of the mean value. }
\end{figure}

\bibliographystyle{apsrev4-1}
\bibliography{ref1}